\documentclass[aps,twocolumn,prd,amssymb,amsmath, bm, floatfix,superscriptaddress,10pt]{revtex4}

\newcommand \beq{\begin{eqnarray}}
\newcommand \eeq{\end{eqnarray}}
\newcommand{\pp}{{\bf p}}
\newcommand{\vv}{{\bf v}}
\newcommand{\qq}{{\bf q}}
\newcommand{\rr}{{\bf r}}
\newcommand{\uu}{{\bf u}}
\newcommand{\jj}{{\bf j}}%\newcommand{\gg}{{\bf g}}

\newcommand{\rme}{{\rm e}}
\newcommand{\rmi}{{\rm i}}

\usepackage{graphicx}
\usepackage{mathptmx}
\begin{document}
\title{Dynamics of the inner crust of neutron stars: hydrodynamics, elasticity and collective modes }
\author{D. Kobyakov}
\affiliation{Department of Physics, Ume{\aa} University, 901 87 Ume{\aa}, Sweden}
\affiliation{Radiophysics Department, Nizhny Novgorod State University, Gagarin Ave. 23, 603950 Nizhny Novgorod, Russia}
\author{C. J. Pethick}
\affiliation{The Niels Bohr International Academy, The Niels Bohr Institute, Blegdamsvej 17, DK-2100 Copenhagen \O, Denmark}
\affiliation{NORDITA, KTH Royal Institute of Technology and Stockholm University, Roslagstullsbacken 23, SE-10691 Stockholm, Sweden}

\begin{abstract}
We present calculations of the hydrodynamics of the inner crust of neutron stars, where a superfluid neutron liquid coexists with a lattice of neutron-rich nuclei.  The long-wavelength collective  oscillations are combinations of phonons in the lattice and phonons in the superfluid neutrons. Velocities of collective modes are calculated from information about effective nucleon--nucleon interactions derived from Lattimer and Swesty's microscopic calculations based on a compressible liquid drop picture of the atomic nuclei and the surrounding neutrons.

\end{abstract}
\maketitle

\section{Introduction}
\label{intro}

Our focus in this paper is on the inner crust of a neutron star which, despite the fact that it occupies rather a small fraction of the total volume of the star, plays an important role in determining observable signals from neutron stars \cite{HaenselPotekhinYakovlev_NeutronStarsI_2007}.   For example, heat from the interior of the star passes through this region on its way to the surface \cite{yakovlev}, torsional oscillations of the crust have been proposed as a mechanism for quasiperiodic oscillations  seen in the afterglows of giant X-ray flares on neutron stars \cite{duncan,watts,andersson}, and the properties of vortex lines in the neutron superfluid have been invoked as a source of glitch phenomena observed in pulsars. Collective modes of the inner crust have been proposed as a means of heat transport \cite{aguilera}.   A pioneering calculation of modes of the inner crust was performed by Epstein \cite{epstein}.
An important physical effect is ``entrainment'', the fact that not all neutrons participate in the motion of the superfluid, since some neutrons are effectively locked to the protons.  This effect is familiar at lower densities in the outer crust, where there are no interstitial neutrons and all neutrons are effectively locked to  the protons.
A general treatment of the hydrodynamics of a neutron superfluid coupled to an elastic solid has been given in Ref.\ \cite{carter}, a macroscopic treatment of entrainment has been given by
\cite{carterChamelHaensel}, a general formulation of the properties of crustal modes have been described in Ref.\ \cite{PethickChamelReddy_ProgrTheorPhys2010}, and  a field-theoretical description of modes has been given in Ref.\ \cite{cirigliano}.
Recently, Chamel, Page and Reddy have calculated frequencies of collective modes of the crust and have investigated some consequences of these modes \cite{ChamelPageReddy}.

In this paper, we consider hydrodynamics and  collective modes of the inner crust. One of our main purposes here is to calculate velocities of long-wavelength collective modes.  As shown in Ref.\ \cite{PethickChamelReddy_ProgrTheorPhys2010}, key ingredients are derivatives of neutron and proton chemical potentials with respect to particle densities.  Calculating them is a challenge, since for the composite system of crystal lattice and superfluid neutrons it is necessary to take into account the equilibrium between neutrons within nuclei and those outside.  In this paper we calculate the derivatives from results of Lattimer and Swesty \cite{lattimer} for thermodynamic quantities; their microscopic calculations of dense matter explicitly allow for the coexistence of nuclei and neutrons.   Another important quantity is the neutron superfluid density, and we demonstrate its strong influence on mode frequencies, which has also been pointed out in Ref.\ \cite{ChamelPageReddy}.

To set the scene, we now give a few basic properties of crustal matter in neutron stars.
It is a very good first approximation to take the temperature to be zero, since it is low compared with the Fermi temperature of the electrons, $\mu_e/k_B$ (where $\mu_e$ is the electron Fermi energy and $k_B$ is the Boltzmann constant) and with the melting temperature
of the lattice for most of the life of a neutron star.   The melting temperature $T_m$ of a Coulomb lattice of ions containing $Z$ protons each is given by
\beq
T_m=\frac1{\Gamma_m}\frac{Z^2e^2}{a_ik_B},
\label{Tm}
\eeq
 where  $a_i=[3/(4\pi n_i)]^{1/3}=(9\pi Z/4)^{1/3}/k_e$ (with $k_e=(3\pi^2n_e)^{1/3}$ being the electron Fermi wave number) is the ion sphere radius and $n_j$, $(j=n,p,e,i)$ are the densities of neutrons, protons, electrons, and ions.  The melting temperature will be affected by the presence of ``dripped'' neutrons in the inner crust, but we expect Eq. (\ref{Tm}) to provide a good first estimate even for this case.  For an electrically neutral plasma consisting of a single species of ion and a rigid background of negative charge, the interaction parameter for melting, $\Gamma_m$ has been computed to be approximately 175 \cite{HaenselPotekhinYakovlev_NeutronStarsI_2007} and therefore
\beq
T_m\approx \left(\frac4{9\pi}\right)^{1/3}\frac1{175}\frac{Z^{5/3}\alpha}{k_B}\hbar k_e c\approx \frac1{100}\left(\frac{Z}{40}\right)^{5/3}\frac{\mu_e}{k_B},
\label{Tm2}
\eeq
where $\alpha=e^2/\hbar c$ is the fine structure constant.
This estimate shows that, given how rapidly neutron stars cool after formation \cite{yakovlev}, the ions are a crystalline solid for most of the life of  the star.

The remainder of this paper is organized as follows.  In Sec.\ \ref{hydro} a description of long-wavelength dynamics is given in terms of hydrodynamics and the theory of elasticity. Expressions for velocities of collective modes are derived in Sec.\ \ref{collective_modes}.   Derivatives of chemical potentials of neutrons and protons with respect to the densities of these components are evaluated in Sec.\ \ref{parameters} from Lattimer and Swesty's microscopic calculations of the properties of matter in the inner crust \cite{lattimer}. There we also comment on the other important parameter determining mode velocities, the neutron superfluid density, and show that scattering from the ionic lattice could be reduced as a result of ionic motion.  We also give a simple expression for the shear elastic constant when the effect of screening by electrons is taken into account.
 In Sec.\ \ref{velocities} we calculate velocities of collective modes and, in particular, show that they are strongly affected by  the magnitude of the neutron superfluid density.
Section \ref{discussion} takes up a number of more detailed topics, and describes outstanding problems and directions for future work.

\section{Hydrodynamic formulation}
\label{hydro}
The inner crust of a neutron star is a three-component system consisting of protons, neutrons and electrons.
The basic long-wavelength modes are plasma oscillations of the electrons and two coupled modes in which the neutrons move in phase or out of phase with the protons and electrons, which are effectively locked together by the Coulomb interaction.  Such modes may be described in terms of the neutron and proton densities, the ion velocity and a quantity specifying the motion of the superfluid neutrons.  The proton, neutron and electron densities vary on length scales comparable to the lattice spacing and the quantities of interest for long-wavelength modes are coarse-grained averages of particle densities taken over distances large compared with the lattice spacing but small compared with the wavelength of the mode.  The superfluid neutrons are characterized by an order parameter corresponding to $\Psi({\bf r})=\langle \psi_\uparrow({\bf r})\psi_\downarrow({\bf r})\rangle$, where $\psi_\sigma({\bf r})$ is the annihilation operator for a neutron with spin $\sigma$ and $\langle \ldots\rangle$ denotes a thermal average.  In the presence of the crystal lattice, $\Psi({\bf r})$ will generally exhibit spatial structure on the scale of the lattice spacing.  For a macroscopic description it is therefore convenient to introduce a phase $\theta({\bf r})$, which is one half of a coarse-grained average of the phase of $\Psi({\bf r})$, the factor of one half being due to the fact that $\Psi({\bf r})$ corresponds to the wave function for a {\it pair} of neutrons with opposite spin.   We then define the neutron superfluid velocity as
\beq
{\bf v}_n^s=\frac{\hbar}m \mathbf{\nabla}\theta.
\label{vs}
\eeq
In this paper we shall extend the standard approach for deriving two-fluid equations for liquid $^4$He \cite{LLhydro} at nonzero temperature and for mixtures of $^3$He in liquid $^4$He \cite{baympethick}.  We shall confine ourselves to long-wavelength phenomena and exclude short length-scale structures such as vortex lines.

We begin by considering flow of neutrons when the ions are at rest.
Phenomenologically, we can express the neutron current density to first order in the superfluid velocity as
\beq
({\bf j}_n)_0=n_n^s ({\bf v}_n^s)_0,
\eeq
and the neutron momentum density is $({\bf g}_n^s)_0=m n_n^s {\bf v}_n^s$.  The quantity $n_n^s$ is the neutron superfluid (number) density, which is related to the neutron superfluid (mass) density $\rho_n^s$ as usually defined by the relationship $\rho_n^s=mn_n^s$
More generally, when nonlinear effects are taken into account, the neutron superfluid density $n_n^s $ will be a function of $v_n^s$ but in the linear theory, which is the subject of this paper, the dependence on $v_n^s$ may be neglected.

Let us now perform a Galilean transformation to a frame in which the velocity of the protons (that of the ionic lattice) is ${\bf v}_p$.  In this frame, the wave function of the neutrons acquires a phase factor \mbox{$\exp({\rmi}m {\bf v}_p\cdot {\bf r}/\hbar)$} at point ${\bf r}$ and therefore the superfluid velocity (\ref{vs}) in the new frame is
\beq
({\bf v}_n^s)=({\bf v}_n^s)_0+{\bf v}_p.
\eeq
  The neutron current in the new frame is given by that in the old frame by the standard result for a Galilean transformation,
\beq
{\bf j}_n= ({\bf j}_n)_0  + n_n {\bf v}_p=    n_n^s ({\bf v}_n^s)_0+n_n {\bf v}_p\nonumber\\=   n_n^s {\bf v}_n^s+n_n^n {\bf v}_p,
\label{neutroncurrent}
\eeq
where $n_n^n=n_n-n_n^s$ is the neutron normal (number) density.  Equation (\ref{neutroncurrent}) has the standard form for the two-fluid model, the velocity of the lattice playing the role of the normal fluid velocity.
Tunneling of protons between different nuclei is unimportant because the proton chemical potential is well below the threshold for proton drip and therefore the current of protons is given by
\beq
{\bf j}_p=n_p {\bf v}_p.
\eeq
In a two-fluid model, the velocity ${\bf v}_p$ corresponds to the velocity of the normal component.

We now give a hydrodynamic description of long-wavelength modes.  We assume that matter is electrically neutral ($n_e=n_p$).  We shall further assume that the inertia of the electrons may be neglected, in which case the electrons respond instantaneously to changes in the proton density.  This is a reasonable approximation, since the effective mass of an electron, $p _e/c$, is small compared with the rest mass of a nucleon.  The basic variables in a two-fluid description are then the densities of neutrons and of protons, the ion velocity, and $\theta$, the phase of the neutron condensate.  We shall assume that, on the timescale of the modes,
weak interaction processes that convert neutrons into protons may be neglected, and therefore the number of neutrons and the number of protons are each conserved.

Conservation of the number of protons is expressed by the equation
\begin{eqnarray}
\label{protonnumber}
{{\partial }_{t}}{{n}_{p}}+\nabla\cdot \left( n_p\mathbf{v}_p \right)=0,
\eeq
and
\beq
\label{neutronnumber}
{{\partial }_{t}}{{n}_{n}}+\nabla \cdot{\bf j}_n={{\partial }_{t}}{{n}_{n}}+\nabla \cdot ( n_{n}^{s}{{\bf v}_{n}} )+\nabla\cdot( n_{n}^{n}{{\bf v}_{p}} )=0.
\eeq
The total momentum density is ${\bf g}={\bf g}_p+{\bf g}_n$ and therefore the condition for momentum conservation is
\beq
{{\partial }_{t}}{\bf g} +\nabla \cdot {\boldsymbol\Pi}=0,
\label{momcons}
\eeq
where $\boldsymbol\Pi$ is the momentum flux density tensor.
The final equation is the Josephson relation, which gives the time evolution of the phase of the condensate.  This reads
\beq
\hbar\partial_t{\theta}+\mu_n=0,
\label{josephson}
\eeq
where $\mu_n$ is the neutron chemical potential.

We now linearize these equations for small deviations of the system from a uniform state with no strain, a stationary lattice, zero neutron superfluid velocity and neutron chemical potential $(\mu_n)_0$.  The continuity equations (\ref{protonnumber}) and (\ref{neutronnumber}) become
\begin{eqnarray}
\label{protonnumberlin}
{{\partial }_{t}}{{n}_{p}}+ n_p\nabla\cdot \mathbf{v}_p =0,
\eeq
and that for neutrons by
\beq
\label{neutronnumberlin}
{{\partial }_{t}}{{n}_{n}}+n_{n}^{s}\nabla \cdot {{\bf v}_{n}} +n_n^n\nabla\cdot{{\bf v}_{p}} =0.
\eeq
In all cases, quantities not acted on by derivatives may be evaluated in the uniform state.

To identify contributions to the momentum flux density tensor it is convenient to consider the energy density of the system for small variations from the uniform state.  The crystal lattice in the ground state is expected to be body centered cubic (bcc), but in a star one expects matter to be polycrystalline, rather than one single crystal.  For wavelengths large compared with the characteristic size of crystallites, the medium behaves as if it were isotropic, in which case the deviation of the energy density $E$ from its value $E_0$ in the equilibrium state is given by
\beq
E-E_0=\frac12\tilde  K u_{ii}^2 + S\left(u_{ij}-\frac{\delta_{ij}u_{kk}}3\right)^2\nonumber \\-L\delta n_nu_{ii} +\frac12\frac{\partial^2E}{\partial n_n^2}(\delta n_n)^2,
\label{Equadratic}
\eeq
where
\beq
u_{ij}=\frac12\left( \frac{\partial u_i}{\partial r_j}+  \frac{\partial u_j}{\partial r_i} \right)
\eeq
and
$L=-{\partial^2E}/{\partial n_n\partial u_{ii}}$.
Changes in the proton density are related to the deformation vector by the relation
\beq
n_p(\rr +\uu)=\frac{n_p(\rr)}{J(\rr+\uu, \rr)},
\label{np}
\eeq
where $J$ is the Jacobian determinant, which for small strains is given by
$J\simeq {(1+\nabla \cdot \uu)}$.
When linearized, Eq.\ (\ref{np}) becomes
\beq
\delta n_p=-n_p\nabla\cdot \uu=-n_pu_{ii}.
\label{deltan_p}
\eeq
It therefore follows that
$L=n_p{\partial^2E}/{\partial n_n\partial n_p}$.  The quantity $\tilde K$ is the contribution to the bulk modulus due to lattice distortions but without changes in the neutron density, and is given by    $\tilde K=n_p^2{\partial^2E}/{\partial  n_p^2}$       and $S$ is the effective shear modulus. The first two terms in Eq.\ (\ref{Equadratic})are the usual expression for the elastic energy \cite[\S 4]{LandauLifshitz-07} of a solid, the third represents the interaction between density deviations of the protons (in nuclei) with those of the neutrons, while the final term is the self-interaction of neutron density deviations.  First order contributions to the change in energy vanish because the initial state is taken to be in equilibrium.

Changes in $\Pi$ come from two sources: one is an isotropic contribution $\delta P\delta_{ik}$ and the second is a shear contribution.  The change in pressure is given by the Gibbs--Duhem relation, which at zero temperature is
\beq
\delta P=n_n \delta \mu_n+n_p\delta \mu_p+n_e\delta \mu_e\\
=n_n \delta \mu_n+n_p\delta \mu_{ep},
\eeq
where the latter expression holds for electrically neutral matter and $\mu_{ep}=\mu_e+\mu_p$ \footnote{In  Ref. \cite{PethickChamelReddy_ProgrTheorPhys2010} the quantity referred to as $\mu_p$ is what we denote in this paper by $\mu_{ep}$.}.
Therefore the changes in $\Pi$ are given by
\beq
\delta \Pi_{ik}=(n_n \delta \mu_n+n_p\delta \mu_{ep})\delta_{ik}- 2S({u}_{ik}-\frac13\delta_{ik}{u}_{ll}).
\eeq
Equation (\ref{momcons}) for momentum conservation then becomes
\beq
m(n_p+n_n^n)\partial_t\vv_p+mn_n^s\partial_t \vv_n^s+n_n\nabla \mu_n +n_p\nabla\mu_{ep}\nonumber\\ -2S\nabla_j({u}_{ij}-\frac13\delta_{ij}{u}_{ll})=0.
\label{momcons2}
\eeq

To linearize the Josephson relation (\ref{josephson}) we introduce the variable $\delta \theta=\theta -(\mu_n)_0~t$, which leads to the equation
\beq
\hbar\partial_t{\delta\theta}+\delta\mu_n=0,
\eeq
or, taking the gradient of this equation,
\beq
m\partial_t{\vv_n^s}+\nabla\delta\mu_n=0.
\label{josephson3}
\eeq
On subtracting $n_n^s$ times this equation from Eq.\ (\ref{momcons2}), one finds
\beq
m(n_p+n_n^n)\partial_t v_{pi}+n_n^n\nabla_i \mu_n +n_p\nabla_i\mu_{ep}\nonumber\\ -2S\nabla_j({u}_{ij}-\frac13\delta_{ij}{u}_{ll})=0,
\label{momcons3}
\eeq
or
\beq
m(n_p+n_n^n)\partial_t \vv_p+n_n^n\nabla\mu_n +n_p\nabla\mu_{ep}\nonumber\\ -\frac43  S\nabla \nabla\cdot \uu +S\nabla\times\nabla\times \uu =0.
\label{momcons3}
\eeq
From Eq.\ (\ref{Equadratic}) one sees that
\beq
\delta \mu_{ep}=-\frac{\tilde K}{n_p}u_{ll} + \frac{L}{n_p}\delta n_n.
\eeq
We show in the Appendix how the equations of motion (\ref{protonnumber}), (\ref{neutronnumber}), (\ref{josephson3}) and (\ref{momcons3}) may be  derived from a variational principle.\\

\section{Collective modes}
\label{collective_modes}
The frequencies and eigenfunctions of long-wavelength modes may be determined by solving Eqs.\ (\ref{protonnumberlin}), (\ref{neutronnumberlin}), (\ref{josephson3}) and (\ref{momcons3}).  We shall assume that physical quantities vary in space and time as $\rme^{\rmi(\qq\cdot \rr-\omega t)}$, where $\qq$ is the wave vector of the mode and $\omega$ its frequency.  Since we take the medium to be isotropic, there is no coupling between longitudinal and transverse modes.

First, we consider transverse modes ($\qq\cdot\uu=0$).  These involve no density changes and the frequencies may be obtained directly from Eq.\  (\ref{momcons3}) and one finds for the velocity $v_t=\omega/q$ of the mode the expression
\beq
v_t^2=\frac{S}{\rho^n},
\label{vt}
\eeq
where $\rho^n=m(n_p+n_n^n)$ is the total normal mass density.  This result, which has previously been pointed out in Refs. \cite{PethickChamelReddy_ProgrTheorPhys2010,ChamelPageReddy}, is the same as for transverse waves in a single-component solid, except that the mass density entering is less than the total mass density $m(n_p+n_n)$ since only the normal neutron density participates in the motion and the neutron superfluid remains stationary.

We turn now to longitudinal modes ($\qq\times\uu=0$).  In Eq.\ (\ref{momcons3}) the last term vanishes and  it is convenient to eliminate $\nabla\cdot \uu$ there by the use of Eq.\ (\ref{deltan_p}).  This gives
\beq
m(n_p+n_n^n)\partial_t \vv_p+n_n^n\nabla\mu_n +n_p\nabla\mu_{ep}\nonumber\\ +\frac43  \frac{S}{n_p}\nabla \delta n_p  =0.
\label{momcons4}
\eeq
Fourier transformation of Eqs.\  (￼\ref{protonnumberlin}), (￼\ref{neutronnumberlin}), (\ref{josephson3}), and (\ref{momcons4}) yields
\begin{widetext}

\begin{equation}
\label{disprellong}
\left(
                     \begin{array}{cccc}
                       v & -n_p & 0&0 \\
                       -(n_n^nE_{np}+n_pE_{pp}+\frac43 {S}/{n_p}) & v\rho^n& -(n_n^nE_{nn}+n_pE_{pn})&0 \\
                       0 & -n_n^n & v&-n_n^s \\
                       -E_{np} &0&-E_{nn}&mv\\
                     \end{array}
                   \right)\left(\begin{array}{c}
  \delta n_p\\
 v_p \\
\delta n_n\\
v_n^s
\end{array}\right)=0.
\end{equation}
\end{widetext}
On eliminating $\delta n_n$ and $\delta n_p$ from this equation, one finds
\begin{equation}
\label{disprellong}
\left(
                     \begin{array}{cc}
                       \rho^n v^2-\mathcal{E}^{nn}-4S/3& -\mathcal{E}^{ns} \\
                       -\mathcal{E}^{ns}& \rho^sv^2-\mathcal{E}^{ss} \\
                                           \end{array}
                   \right)\left(\begin{array}{c}
  v_p \\
v_n^s
\end{array}\right)=0,
\end{equation}
where
\beq
\mathcal{E}^{kl}&=&n_i^kn_j^lE_{ij}\nonumber\\&=&E_{pp}n_p^kn_p^l+E_{pn}(n_p^kn_n^l+n_n^kn_p^l)+E_{nn}(n_n^n)^2.
\label{calE}
\eeq
The upper indices $k$ and $l$ refer to the normal ($n$) and superfluid ($s$) components and  $\rho^s=mn_n^s$ is the neutron superfluid mass density. For the protons, which are normal, $n_p^n=n_p$ and \mbox{$n_p^s=0$.}

To understand the physics of the modes, it is convenient to find the frequencies of modes when the lattice is forced to remain stationary, and when the superfluid velocity is forced to be zero.    The velocity of the mode for a stationary lattice is given by
\beq
v_n^2=\frac{\mathcal{E}^{ss}}{\rho^s}=\frac{n_n^sE_{nn}}{m}.
\label{vn}
\eeq
This mode corresponds to a phonon in the neutron superfluid (the Bogoliubov--Anderson mode) and its velocity is given by the same expression as that for phonons in a  Bose--Einstein condensate in an optical lattice \cite{kramer}.  The velocity of the lattice modes when the superfluid is stationary is given by
\beq
v_p^2=\frac{\mathcal{E}^{nn}+\frac43 S}{\rho^n}.
\label{vp}
\eeq
The ``force constant'' for the mode depends on the energy due to changes in the densities proportional to the normal densities of the two species,  together with an additional contribution due to the shearing of the lattice. The effective mass density is the total normal mass density.
This mode corresponds to an oscillation of the lattice and entrained neutrons but no motion of the superfluid neutrons.  In general, the two modes are coupled and their velocities $v=\omega/q$ are given by
\beq
(v^2-v_n^2)(v^2-v_p^2)-   v_{np}^4=0,
\label{modevel}
\eeq
where
\beq
 v_{np}^2=\frac{\mathcal{E}^{ns}}{\sqrt{\rho^s\rho^n}}=\left(\frac{n_n^s}{m\rho^n}\right)^{1/2}    (E_{nn}n_n^n+E_{np}n_p).
 \label{vnp}
\eeq
The mode frequencies are therefore given by
\beq
v_{\pm}^2=\frac{v_n^2+v_p^2}2 \pm\sqrt{\left(\frac{v_n^2-v_p^2}{2}\right)^2+  v_{np}^4}\,\,\,,
\label{vpm}
\eeq
or
\beq
v_{\pm}^2=\frac{v_n^2+v_p^2}2 \pm\sqrt{\left(\frac{v_n^2+v_p^2}{2}\right)^2+  v_{np}^4-v_n^2v_p^2}\,\,\,.
\eeq
Since
\beq
 v_{np}^4-v_n^2v_p^2=\frac{n_p^2(n_n^s)^2}{\rho^s\rho^n}\left(E_{nn}[E_{pp}+4S/3]-E_{np}^2\right),
\eeq
one sees that, if the shear modulus is neglected, one of the mode velocities will become imaginary if one of the conditions for thermodynamic stability to long-wavelength variations of the neutron and proton densities,
$\det E_{ij}>0$, is violated.

The way in which we have written the equation for the mode velocities differs from that which is used in Ref.\ \cite{ChamelPageReddy}.  We have chosen to include all effects of entrainment, which gives rise to a current-current (vector) interaction, in the velocities $v_n$ and $v_p$. As a consequence only a scalar (density-density) coupling between these modes remains in Eq.\ (\ref{modevel}).  In Ref.\ \cite{ChamelPageReddy}, $v_\phi$ corresponds to our $v_n$ but their lattice phonon mode is different from our mode with velocity $v_p$ and the coupling between their two modes has a vector character.

\section{Evaluation of parameters}
\label{parameters}

The basic assumption made in the hydrodynamic treatment is that matter remains electrically neutral. This is a good approximation provided the wave number is small compared with the  Thomas-Fermi screening wave number  of the electrons
\begin{equation}
\label{k_TF}
k_{TF}=\sqrt{4\pi e^2 dn_e/d\mu_e}\simeq \left(\frac{4\alpha}{\pi}\right)^{1/2}k_e.
\end{equation}
Here $n_e$ is the electron density, $\mu_e$ the electron chemical potential, and $k_e$ is the electron Fermi momentum.   The second expression in Eq.\ (\ref{k_TF}) holds for ultrarelativistic electrons.  The condition for electrical neutrality at long wavelengths holds provided the frequency is much less than the electron plasma frequency
 \beq
\omega_{pe}=\left(\frac{4\pi n_e e^2 }{m_e}\right)^{1/2} \left(\frac{1}{\hbar^2 k_e^2/m_e^2c^2+1}\right)^{1/4}\simeq  \left(\frac{4\alpha}{3\pi}\right)^{1/2}k_e c  .\nonumber \\
\eeq

For given proton and neutron densities, the mode frequencies depend on the parameters $E_{nn}, E_{pp}, E_{np}$, the shear elastic constant, and the neutron superfluid density $n_n^s$, which we now discuss in turn.

\subsection{E$_{nn}$, E$_{np}$, and E$_{pp}$}

Determination of the $E_{ij}$ for the composite system of nuclei immersed in a neutron gas requires information about the equilibrium state of this system.  This may be determined from microscopic calculations.  We have chosen to use the work of Lattimer and Swesty \cite{lattimer}, which is based on Ref.\ \cite{LPRL}.  One reason for doing this is that the model takes into account equilibrium between neutrons in nuclei and those outside and uses nucleon--nucleon interactions that have been fitted to masses of laboratory nuclei and the properties of pure neutron matter.  A second reason is that second derivatives are evaluated analytically without resort to finite differences, thereby reducing numerical noise.  A third reason is
that it is one of the models of dense matter most widely used in simulations.  A fourth is that extensive tables of physical quantities for the model are available.  The particular variant of the model is relatively unimportant at subnuclear densities, and the one we used corresponds to the parameters $S_v=29.35$ MeV, $K=370.6$ MeV, $n_s=0.155$ fm$^{-3}$, $B=16$ MeV, $m^*=0.911$ , $m^*_n=1.064 m_n$ and $S_v'=11.77$ MeV and
$L=3S_v'$.

The general formalism described earlier applies for arbitrary neutron excess but one expects that in the inner crust of neutron stars matter is close to being in equilibrium with respect to weak processes and consequently we describe results for this case.   The condition for weak equilibrium is that the energy to add a neutron be the same as the energy to add a proton and an electron, i.e.,
\beq
\mu_n=\mu_p+\mu_e=\mu_{ep}.
\eeq
This condition is equivalent to
\beq
{\tilde \mu}\equiv \mu_n-\mu_p-\mu_e=-\frac1n\frac{\partial E}{\partial Y_e}=0,
\eeq
where $n=n_p+n_n$ is the total baryon density and $Y_e=n_p/n$ is the fractional proton concentration.
While $\tilde \mu=0$ in the unperturbed state, it is nonzero during an oscillation because weak equilibrium cannot be established on the timescale of the modes.
We took a temperature of $0.01$ MeV, which is small compared with typical nuclear energy scales.

On the Lattimer--Swesty website one can find values of $\partial p/\partial n=\partial^2E/\partial n^2$ and $\partial p/\partial Y_e$.  In beta equilibrium,
\beq
\frac{\partial p}{\partial Y_e}=n\frac{\partial^2 E }{\partial n\partial Y_e }.
\eeq

To determine all the $E_{ij}$ we need a third quantity, and James Lattimer kindly prepared for us a table of $\partial{\hat \mu}/\partial Y_e$, where ${\hat \mu=\mu_n-\mu_p}$, from which one can calculate the quantity
\beq
\frac{\partial{\tilde \mu}}{\partial Y_e}= \frac{\partial{\hat \mu}}{\partial Y_e}-n\frac{\partial \mu_e}{\partial Y_e}=\frac1n\frac{\partial^2 E }{\partial Y_e^2 },
\eeq
where the second expression holds in beta equilibrium.   The calculations of Lattimer and Swesty are for electrically neutral matter and therefore from their results it is possible to determine the sum of the proton and electron chemical potentials but not each chemical potential separately.  In their approach, the energy is expressed as the sum of two contributions, the first being that of a gas of noninteracting electrons and the second being the rest, which includes the effects of nucleon--nucleon interactions and the Coulomb interaction.  What is given as the electron chemical potential is the value for a free Fermi gas,
\beq
\mu_e=(p_e^2c^2 +m_e^2c^4)^{1/2}\approx \hbar c (3\pi^2n_e)^{1/3},
\eeq
and all contributions from the Coulomb interaction are included in $\mu_p$.  More generally, if one wishes to allow for the electron and proton densities to be unequal, the electron chemical potential should include the contribution $-e$ times the spatial average of the electrostatic potential.  In the present paper $\mu_p$ and $\mu_e$ enter only as their sum, and therefore it does not matter how Coulomb contributions are apportioned between the two chemical potentials.

The calculations of Lattimer and Swesty give second derivatives of the energy density with respect to the variables $n$ and $Y_e$, but the quantities we need are derivatives with respect to $n_n$ and $n_p$.  By changing variables, one finds that the two sets of quantities are related by the transformation
\begin{equation}
\label{matrix_formula}
\left(\begin{array}{c}
  E_{nn} \\
  E_{np} \\
  E_{pp}
\end{array}\right)={1\over{n}}\left(
                     \begin{array}{ccc}
                       1 & -2Y_e & -Y_e^2 \\
                       1 & 1-2Y_e & Y_e(1-Y_e) \\
                       1 & 2(1-Y_e) & -(1-Y_e)^2 \\
                     \end{array}
                   \right)\left(\begin{array}{c}
  dp/dn \\
  n^{-1}dp/dY_e \\
  d\tilde{\mu}/dY_e
\end{array}\right).
\end{equation}

  In Table I we exhibit results for physical properties of crustal matter.  For the mass number $A$ we have taken the number of nucleons within the volume of the nucleus plus the number of surface neutrons, and $w$ is the fractional volume occupied by nuclei. One noteworthy result is that $E_{pp}$ is close to the contribution $\partial \mu_e/\partial n_e$ from a uniform electron gas at essentially all densities.  The Coulomb lattice contribution is of relative order $Z^{2/3}\alpha$, is negative, and is always small in magnitude compared with $\partial \mu_e/\partial n_e$.  In addition, contributions to $E_{pp}$ from nuclear forces are numerically small but they are positive and overwhelm the lattice contribution at higher densities. The effective neutron--proton interaction $E_{np}$, which is attractive, is appreciable.

\subsection{Shear elastic constant}
In high density matter, there are a number of contributions to the elastic constants.  The first is the Coulomb energy due to interactions between nuclei.  This contribution was calculated for a rigid background of electrons in Ref.\ {\cite{shearFuchs1936}} and it is of order $n_IZ^2e^2/a_i$.  A second contribution is due to the non-uniformity of the electron density and this is negative and of order $k_{TF}^2 a_i^2$ times the result for a rigid electron background. As we remarked earlier, we assume that the medium is effectively isotropic.    To relate the effective shear elastic constant of the polycrystal to those of the single crystal is a difficult problem, and a number of suggestions have been made \cite{hirsekorn}.  In an astrophysical context,  Ogata and Ichimaru \cite{shearNS_OgataIchimaru_1990} did this by averaging elastic constants over all possible directions, a procedure which Baiko adopted in his recent work \cite{baiko}.  For a static lattice of ions and a rigid background of electrons  Ogata and Ichimaru find
\beq
S_{\rm Coul}=0.1194n_i\frac{Z^2e^2}{a_i}
\eeq
and  the contribution due to inhomogeneity of the electron density is well fit for $Z\lesssim 40$ by
\beq
S_{\rm screen}=-0.010Z^{2/3}[1+(m_e c/p_e)^2]^{1/2}S^{\rm Coul}
\eeq
For applications to the inner crust of neutron stars, electrons may be treated as ultrarelativistic.
The shear modulus has a negative contribution from lattice phonons and this has been calculated in Ref.\ \cite{Baiko2011} in the absence of dripped neutrons. We find that for $T\sim10^8K$, the phonon contribution is about 2 orders of magnitude smaller than the contributions for a static lattice and consequently we neglect it.
The total shear elastic constant is given by
\beq
 S&=&S_{\rm Coul}+S_{\rm screen}\nonumber\\
 &=&0.1194 \, n_i\frac{Z^2e^2}{a_i}\left(1-0.010Z^{2/3}\right),
\eeq
where in the last term we have taken the ultrarelativistic limit.

\subsection{Neutron superfluid density}

A key quantity is the neutron superfluid density, $n_n^s$.  When nuclei occupy a small fraction of space, one would expect on physical grounds that this would be close to the density of neutrons outside nuclei, which we denote by $n^{\rm out}$.   However, in a recent series of papers, Chamel has carried out Hartree--Fock calculations of the band structure of neutrons in the inner crust of neutron stars  which indicate that $n_n^s$ could be as low as $n^{\rm out}/10$ or less at matter densities 1/10 to 1/3 of nuclear matter density \cite{Chamel_BandStructCalcul_NS_2012}.  There are a number of physical effects, such the effect of lattice vibration on the band structure, which we shall consider later in this subsection,  and the influence of disorder and impurities that need to be investigated in order to reduce the uncertainly in $n_n^s$. To examine the sensitivity of collective mode properties to $n_n^s$ we have carried out two sets of calculations, one with $n_n^s=n^{\rm out}$ and the other with $n_n^s$ as a function of density given by the results of Chamel's calculations.

The calculations in Ref.\ \cite{Chamel_BandStructCalcul_NS_2012} were carried out for a rigid lattice of nuclei and one may ask whether the motion of ions could affect this.   In the harmonic approximation for the lattice and for ions that have no internal degrees of freedom, the strength of the  lattice potential corresponding to a lattice wave vector $\bf G$ is reduced by a factor $\exp[-W({\bf G})]$, which is the square root of the Debye--Waller factor as conventionally defined \cite{Kittel_QuantumTheoryOfSolids}.  For a cubic crystal,
\beq
W(G)= \frac16 G^2 \langle u^2\rangle,
\eeq
where $\langle u^2\rangle$ is the mean square displacement of an ion. At zero temperature, for a bcc lattice of ions and a rigid background of electrons, one finds in the harmonic approximation that
\begin{figure}
\includegraphics[width=3.75in]{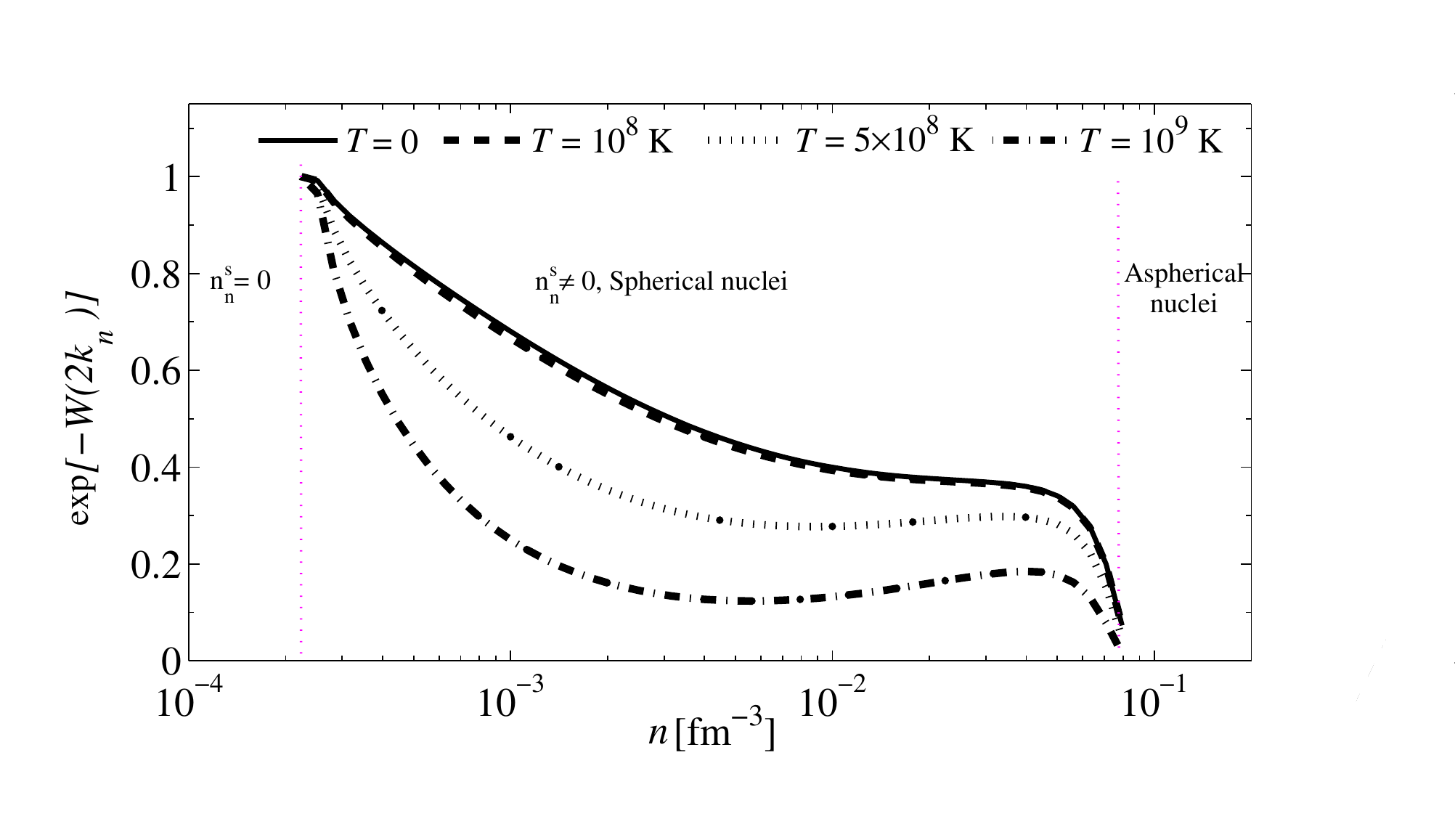}
\caption{The factor $\exp(-W(2k_n^{\rm out}))$ by which the scattering of the periodic lattice is reduced by lattice vibrations for four different temperatures.  The quantity is calculated for the maximum momentum transfer $2k_n^{\rm out}$ of free neutrons at the Fermi surface for a number density equal to $n_n^{\rm out}$ and is therefore the minimum value of the factor.}
\end{figure}

\beq
\label{w_0}
 \langle u^2\rangle=\frac{\hbar}{Vmn_i}{\sum_{\mathbf{k},\epsilon}}
\frac{1}{\omega_{\mathbf{k},\epsilon}}
\left(\frac12 +n_{\mathbf{k},\epsilon}\right),
\eeq
where $\omega_{\mathbf{k},\epsilon}$ is the frequency of a phonon with wave vector $\mathbf{k}$ and polarization $\epsilon$, and $n_{\mathbf{k},\epsilon}=1/[\exp(\hbar \omega_{\mathbf{k},\epsilon}/k_BT)-1]$ is the Bose distribution function.
Thus one finds for zero temperature that \cite{PRE64.057402.Baiko2001b}
\beq
\langle u^2\rangle_{T=0}=\frac32\frac{\hbar u_{-1}}{ m_i \omega_{pi}},
\eeq
where $u_{-1}\approx 2.79855 $ is the average of the inverse of the phonon frequency over the Brillouin zone, in units of $1/\omega_{pi}$.  Here the ion plasma frequency  $\omega_{pi}$  is given by
\beq
\omega_{pi}^2=\frac{4\pi Z^2e^2 n_i}{m_i},
\eeq
where $m_i$ is the mass of an ion.  At the beginning of this subsection we pointed to the uncertainty in the value of the neutron superfluid density.  This leads to a corresponding uncertainty in the  effective mass of a nucleus, which is $\rho^n/n_i=m_p(n-n_n^s)/n_i$.  So as not to overestimate $\langle u^2\rangle$ we take it to be $Am_p$.
For nonzero temperature, one may estimate the thermal contribution to $\langle u^2\rangle$ by using the Debye model for the transverse modes, which are taken to have a constant velocity,  and putting all longitudinal modes at a common frequency.  One then finds
\begin{align}
\langle u^2\rangle=\frac{\hbar }{ m_i \omega_{pi}}
\left(\frac32 u_{-1} +\frac{6}{\alpha_t^2} \frac{T}{T_{pi}} D_1(\alpha_tT_{pi}/T)   \right.\nonumber \\
\left.+\frac1{\alpha_l}\frac{1}{\exp(\alpha_lT_{pi}/T)-1}     \right),
\label{u2T}
\end{align}
where $D_1(x)=x^{-1}\int_0^x dy {y/({e^y-1})}$. By fitting the model to the second moment of the phonon frequencies and to the average of the logarithm of the phonon frequencies for a bcc crystal with a rigid background of negative charge, one finds $\alpha_{t}=0.426548$ and $\alpha_{l}=0.88412$  \cite[p.\ 82]{HaenselPotekhinYakovlev_NeutronStarsI_2007}. The expression (\ref{u2T}) is a good approximation for all temperatures below the melting temperature.  In Fig. 1 we present numerical calculation of $\rme^{-W(2k_n^{\rm out})}$ for a number of temperatures.  The momentum transfer $2k_n^{\rm out}=2(3\pi^2n_n^{\rm out})^{1/3}$ is the maximum value of the momentum transfer for neutrons on a spherical Fermi surface, and therefore this calculation gives a lower bound on the value of $\rme^{-W}$ for other relevant momentum transfers.  This calculation indicates that the Debye--Waller factor can be important in reducing scattering from the periodic lattice and its effect on neutron band structure and the neutron superfluid density should be investigated in greater detail.

\begingroup
\squeezetable
\begin{table*}
\caption{Microscopic parameters of crustal matter. The quantity $x^{\rm in}$ is the proton fraction of matter in the interior of nuclei.}
\label{tab1}
\begin{center}
\begin{tabular}{|c|c|c|c|c|c|c|c|c|c|c|c|}
  \hline
  $n [\texttt{fm}^{-3} ]$&$n^{\rm out}  [\texttt{fm}^{-3} ]$&$\mu_e  [\texttt{MeV}]$&$Y_e=n_p/n$&$x^{\rm in}$ &$w$&$A$&$Z$&$E_{nn}\,[\texttt{MeV}\,\texttt{fm}^3] $&$E_{np}\,[\texttt{MeV}\,\texttt{fm}^3] $&$E_{pp}\,[\texttt{MeV}\,\texttt{fm}^3] $&$\partial\mu_e/\partial n_e \,[\texttt{MeV}\,\texttt{fm}^3]$\\ \hline

2.512E-04&1.138E-06&2.729E+01&3.557E-01&3.573E-01&1.380E-03&106.0&37.9&3.937E+04&-4.281E+04&1.061E+05&1.018E+05\\
2.818E-04&1.877E-05&2.772E+01&3.320E-01&3.557E-01&1.451E-03&107.0&38.1&2.395E+04&-2.236E+04&7.081E+04&9.870E+04\\
3.162E-04&4.239E-05&2.806E+01&3.070E-01&3.544E-01&1.510E-03&107.8&38.2&1.906E+04&-1.963E+04&6.772E+04&9.632E+04\\
3.548E-04&7.051E-05&2.838E+01&2.832E-01&3.532E-01&1.567E-03&108.5&38.3&1.551E+04&-1.800E+04&6.685E+04&9.414E+04\\
3.981E-04&1.032E-04&2.870E+01&2.610E-01&3.521E-01&1.626E-03&109.3&38.5&1.274E+04&-1.668E+04&6.664E+04&9.206E+04\\
4.467E-04&1.407E-04&2.903E+01&2.405E-01&3.509E-01&1.686E-03&110.0&38.6&1.054E+04&-1.548E+04&6.667E+04&9.002E+04\\
5.012E-04&1.836E-04&2.936E+01&2.218E-01&3.496E-01&1.750E-03&110.8&38.7&8.757E+03&-1.435E+04&6.672E+04&8.800E+04\\
5.623E-04&2.323E-04&2.970E+01&2.047E-01&3.484E-01&1.818E-03&111.6&38.9&7.315E+03&-1.327E+04&6.672E+04&8.598E+04\\
6.310E-04&2.877E-04&3.006E+01&1.891E-01&3.470E-01&1.891E-03&112.5&39.0&6.141E+03&-1.224E+04&6.660E+04&8.395E+04\\
7.079E-04&3.504E-04&3.043E+01&1.749E-01&3.457E-01&1.969E-03&113.4&39.2&5.181E+03&-1.127E+04&6.634E+04&8.190E+04\\
7.943E-04&4.213E-04&3.083E+01&1.620E-01&3.442E-01&2.054E-03&114.4&39.4&4.394E+03&-1.035E+04&6.592E+04&7.982E+04\\
8.913E-04&5.014E-04&3.124E+01&1.503E-01&3.427E-01&2.147E-03&115.4&39.5&3.746E+03&-9.491E+03&6.534E+04&7.772E+04\\
1.000E-03&5.918E-04&3.168E+01&1.397E-01&3.411E-01&2.248E-03&116.5&39.7&3.211E+03&-8.688E+03&6.461E+04&7.558E+04\\
1.122E-03&6.938E-04&3.214E+01&1.300E-01&3.394E-01&2.359E-03&117.7&39.9&2.766E+03&-7.942E+03&6.372E+04&7.342E+04\\
1.259E-03&8.087E-04&3.263E+01&1.213E-01&3.376E-01&2.480E-03&118.9&40.1&2.397E+03&-7.253E+03&6.268E+04&7.123E+04\\
1.413E-03&9.382E-04&3.315E+01&1.133E-01&3.357E-01&2.614E-03&120.2&40.4&2.087E+03&-6.618E+03&6.151E+04&6.902E+04\\
1.585E-03&1.084E-03&3.370E+01&1.061E-01&3.336E-01&2.762E-03&121.7&40.6&1.827E+03&-6.036E+03&6.021E+04&6.678E+04\\
1.778E-03&1.248E-03&3.428E+01&9.957E-02&3.315E-01&2.926E-03&123.2&40.8&1.607E+03&-5.502E+03&5.879E+04&6.453E+04\\
1.995E-03&1.432E-03&3.490E+01&9.365E-02&3.292E-01&3.107E-03&124.9&41.1&1.421E+03&-5.015E+03&5.727E+04&6.225E+04\\
2.239E-03&1.639E-03&3.556E+01&8.827E-02&3.268E-01&3.309E-03&126.7&41.4&1.262E+03&-4.571E+03&5.567E+04&5.997E+04\\
2.512E-03&1.872E-03&3.626E+01&8.340E-02&3.243E-01&3.534E-03&128.6&41.7&1.125E+03&-4.166E+03&5.398E+04&5.768E+04\\
2.818E-03&2.134E-03&3.700E+01&7.899E-02&3.216E-01&3.785E-03&130.7&42.0&1.007E+03&-3.798E+03&5.223E+04&5.539E+04\\
3.162E-03&2.428E-03&3.779E+01&7.499E-02&3.187E-01&4.066E-03&132.9&42.4&9.045E+02&-3.464E+03&5.043E+04&5.311E+04\\
3.548E-03&2.758E-03&3.862E+01&7.136E-02&3.156E-01&4.382E-03&135.3&42.7&8.150E+02&-3.161E+03&4.858E+04&5.084E+04\\
3.981E-03&3.128E-03&3.951E+01&6.807E-02&3.124E-01&4.737E-03&137.9&43.1&7.363E+02&-2.886E+03&4.671E+04&4.859E+04\\
4.467E-03&3.544E-03&4.045E+01&6.509E-02&3.090E-01&5.137E-03&140.8&43.5&6.667E+02&-2.637E+03&4.482E+04&4.636E+04\\
5.012E-03&4.010E-03&4.144E+01&6.239E-02&3.054E-01&5.589E-03&143.8&43.9&6.049E+02&-2.412E+03&4.292E+04&4.417E+04\\
5.623E-03&4.533E-03&4.249E+01&5.994E-02&3.016E-01&6.101E-03&147.1&44.4&5.496E+02&-2.207E+03&4.102E+04&4.201E+04\\
6.310E-03&5.120E-03&4.360E+01&5.772E-02&2.975E-01&6.681E-03&150.7&44.8&5.000E+02&-2.021E+03&3.914E+04&3.990E+04\\
7.079E-03&5.777E-03&4.477E+01&5.571E-02&2.933E-01&7.340E-03&154.6&45.3&4.553E+02&-1.853E+03&3.727E+04&3.784E+04\\
7.943E-03&6.514E-03&4.601E+01&5.388E-02&2.888E-01&8.092E-03&158.8&45.9&4.148E+02&-1.700E+03&3.544E+04&3.583E+04\\
8.913E-03&7.340E-03&4.731E+01&5.221E-02&2.841E-01&8.949E-03&163.4&46.4&3.780E+02&-1.561E+03&3.365E+04&3.389E+04\\
1.000E-02&8.266E-03&4.868E+01&5.069E-02&2.791E-01&9.931E-03&168.4&47.0&3.445E+02&-1.435E+03&3.190E+04&3.201E+04\\
1.122E-02&9.303E-03&5.011E+01&4.930E-02&2.738E-01&1.106E-02&173.7&47.6&3.139E+02&-1.320E+03&3.020E+04&3.020E+04\\
1.259E-02&1.047E-02&5.162E+01&4.802E-02&2.683E-01&1.235E-02&179.5&48.2&2.860E+02&-1.216E+03&2.855E+04&2.846E+04\\
1.413E-02&1.177E-02&5.320E+01&4.684E-02&2.626E-01&1.384E-02&185.8&48.8&2.604E+02&-1.120E+03&2.697E+04&2.680E+04\\
1.585E-02&1.323E-02&5.485E+01&4.575E-02&2.565E-01&1.556E-02&192.7&49.4&2.370E+02&-1.033E+03&2.545E+04&2.521E+04\\
1.778E-02&1.486E-02&5.657E+01&4.473E-02&2.502E-01&1.755E-02&200.0&50.0&2.156E+02&-9.524E+02&2.399E+04&2.370E+04\\
1.995E-02&1.670E-02&5.836E+01&4.378E-02&2.435E-01&1.987E-02&207.9&50.6&1.962E+02&-8.789E+02&2.260E+04&2.227E+04\\
2.239E-02&1.875E-02&6.023E+01&4.289E-02&2.366E-01&2.257E-02&216.4&51.2&1.785E+02&-8.114E+02&2.128E+04&2.091E+04\\
2.512E-02&2.106E-02&6.217E+01&4.204E-02&2.293E-01&2.575E-02&225.5&51.7&1.626E+02&-7.493E+02&2.003E+04&1.962E+04\\
2.818E-02&2.364E-02&6.419E+01&4.124E-02&2.216E-01&2.950E-02&235.0&52.1&1.484E+02&-6.923E+02&1.885E+04&1.841E+04\\
3.162E-02&2.653E-02&6.629E+01&4.049E-02&2.136E-01&3.397E-02&244.9&52.3&1.358E+02&-6.399E+02&1.774E+04&1.726E+04\\
3.548E-02&2.977E-02&6.849E+01&3.980E-02&2.051E-01&3.936E-02&254.9&52.3&1.249E+02&-5.919E+02&1.670E+04&1.617E+04\\
3.981E-02&3.339E-02&7.079E+01&3.917E-02&1.960E-01&4.596E-02&264.4&51.8&1.156E+02&-5.481E+02&1.573E+04&1.513E+04\\
4.467E-02&3.743E-02&7.322E+01&3.863E-02&1.862E-01&5.421E-02&272.6&50.8&1.080E+02&-5.083E+02&1.483E+04&1.414E+04\\
5.012E-02&4.193E-02&7.581E+01&3.820E-02&1.756E-01&6.480E-02&277.4&48.7&1.021E+02&-4.724E+02&1.399E+04&1.320E+04\\
5.623E-02&4.693E-02&7.857E+01&3.791E-02&1.639E-01&7.884E-02&275.3&45.1&9.788E+01&-4.406E+02&1.322E+04&1.228E+04\\
6.310E-02&5.244E-02&8.155E+01&3.777E-02&1.507E-01&9.828E-02&259.8&39.2&9.532E+01&-4.126E+02&1.253E+04&1.140E+04\\
7.079E-02&5.849E-02&8.476E+01&3.780E-02&1.357E-01&1.267E-01&220.2&29.9&9.446E+01&-3.883E+02&1.190E+04&1.056E+04\\
7.943E-02&6.510E-02&8.819E+01&3.795E-02&1.183E-01&1.712E-01&145.6&17.2&9.562E+01&-3.665E+02&1.135E+04&9.751E+03\\
8.913E-02&7.241E-02&9.178E+01&3.813E-02&9.828E-02&2.466E-01&45.7&4.5&1.011E+02&-3.429E+02&1.086E+04&9.003E+03\\

 \hline
\end{tabular}
\end{center}
\end{table*}
\endgroup

\section{Mode velocities}
\label{velocities}

The transverse and longitudinal sound velocities (Eqs.\ (\ref{vt}) and (\ref{vpm})) are plotted in Fig 1. Velocities of longitudinal modes  are indicated by solid lines,  and dashed lines show the  velocities of the uncoupled modes, $v_n$ and $v_p$, given by Eqs.\ (\ref{vn}) and (\ref{vp}) .  As one sees, coupling between modes is relatively weak.  With increasing normal neutron density, the velocities of both modes are reduced:  a higher neutron normal density decreases the velocity of lattice phonon, while a decreasing neutron superfluid density decreases the velocity of sound in the neutron superfluid.

In Fig. 2 we show a magnified plot of the longitudinal velocities for $n_n^s=n_n^{\rm out}$ for a range of  densities around 6$\times 10^{-4}$ fm$^{-3}$ where the velocities of the two modes are close to each other.  The coupling between the two uncoupled modes,  Eq.\ (\ref{vnp}), passes through zero because the $E_{nn}$  and $E_{np}$ terms  cancel.

\begin{figure}
\includegraphics[width=3.75in]{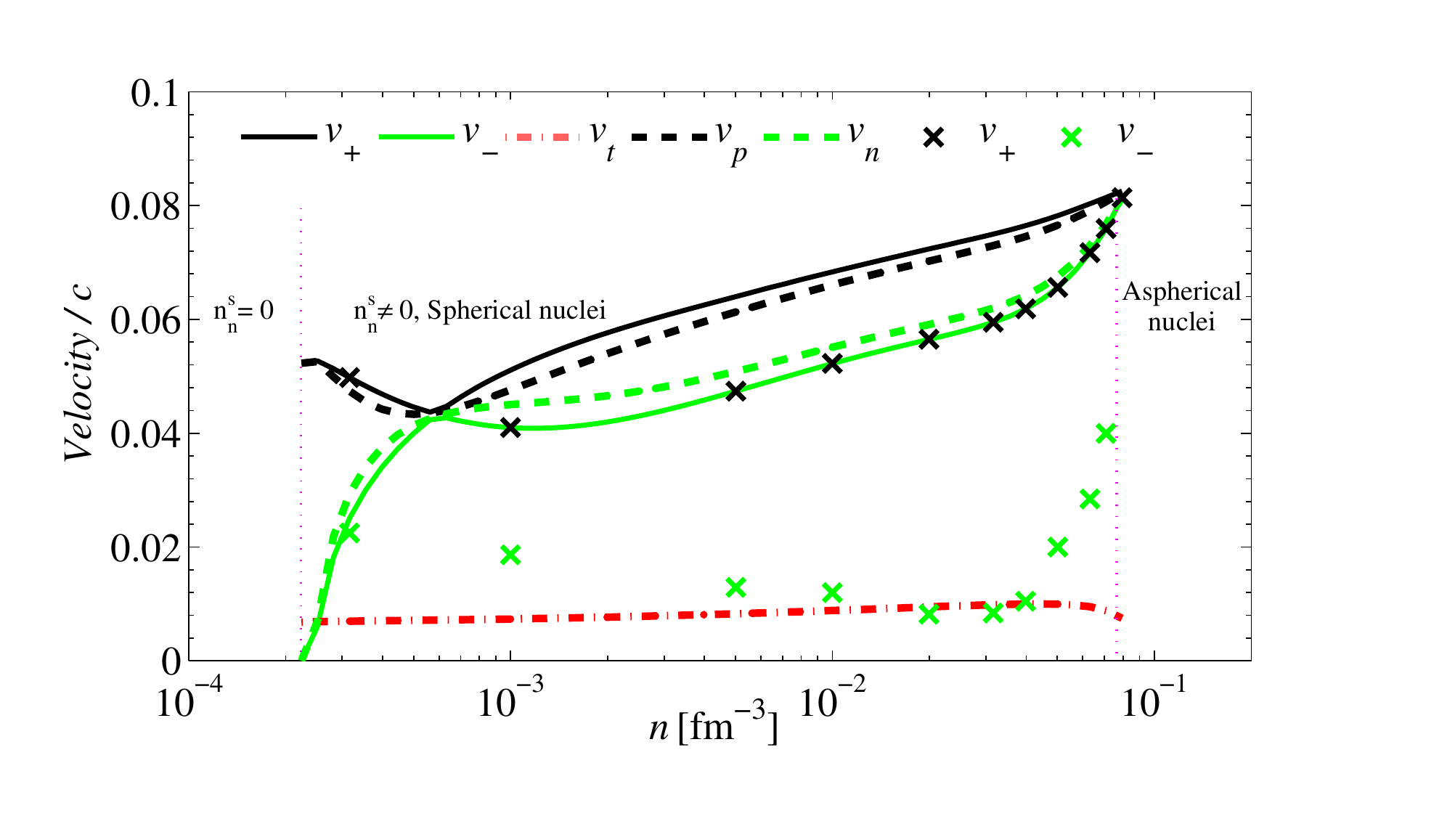}
\caption{(Color online) Velocities of long-wavelength modes as a function of nucleon density.  Velocities of uncoupled modes are given by $v_{n}$ and $v_p$, the corresponding velocities of the coupled modes by $v_\pm$, and the velocity of the transverse mode by  $v_{t}$. Curves are for $n_n^s=n^{\rm out}$ while the crosses show results for $n_n^s$ taken from Chamel's calculation \cite{Chamel_BandStructCalcul_NS_2012}.}
\end{figure}

\begin{figure}
\includegraphics[width=3.75in]{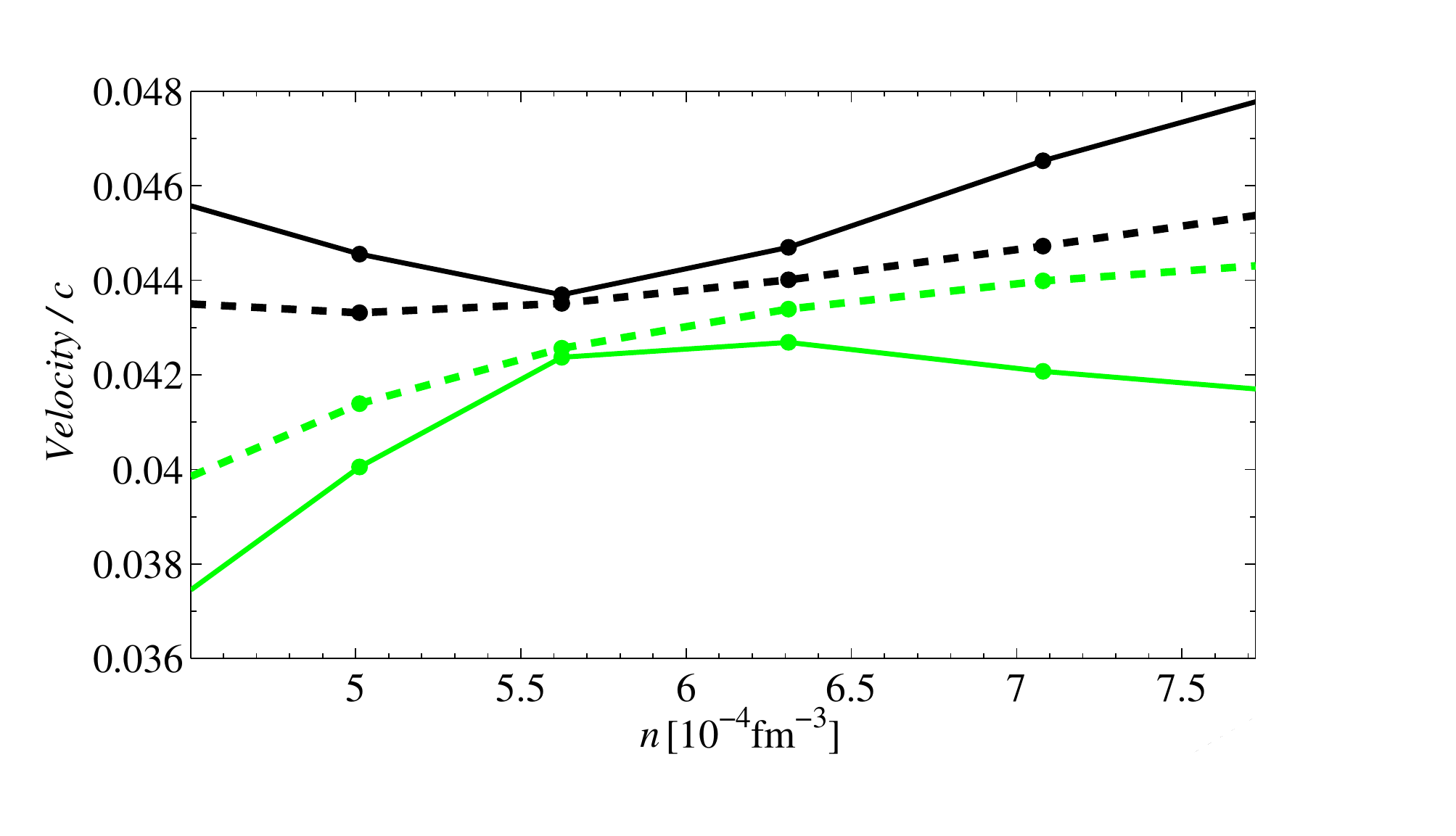}
\caption{(Color online) Expanded version of the plot in Fig. 2 for densities at which the longitudinal mode velocities approach each other most closely for $n_n^s=n^{\rm out}$ (see text).  The dots show computed values, which have been joined by straight lines.}
\end{figure}

The calculations demonstrate that the reduction of the neutron superfluid density predicted by Chamel would have a dramatic effect on both longitudinal modes, whose velocities are reduced compared with the values for $n_n^s=n^{\rm out}$, as has also been pointed out by Chamel et al. \cite{ChamelPageReddy}.  For the Chamel results for $n_n^s$ the velocities we find are in reasonable agreement with the calculations of Ref.\ \cite{ChamelPageReddy}.

\section{Discussion and concluding remarks}
\label{discussion}

In the general formalism, we used as the basic variables the lattice displacement, $\uu$, and the phase of the condensate, $\theta$, in addition to the neutron and proton densities.  These are the ones traditionally used in condensed matter physics, and they have the advantage that the kinetic energy  is a sum of a term proportional to ${\dot u}^2$ and one proportional to $(\nabla \theta)^2$, with no cross terms (see Eq.\ (\ref{A4})).  This is similar to what is done in discussing stability of two-component systems, when it is convenient to work in terms of the density of one component and the chemical potential of the other one \cite{callen}.  This approach brings out clearly from the outset the fact that there is just \emph{one} additional parameter needed to describe the kinetic energy of the coupled superfluid--lattice system compared with the uncoupled system.

In calculating thermodynamic derivatives we used the calculations of Lattimer and Swesty, which treat $Z$ as a continuous variable, which depends on density and proton fraction.  Physically, it would be more realistic to keep $Z$ fixed,  since the time scales for changing $Z$, which requires either weak interactions or major rearrangements of the ionic lattice, are very long compared with periods of typical oscillations of the crust.   However, in calculating derivatives of thermodynamic quantities we have allowed $Z$ to assume its equilibrium value, but, since the variations of $Z$ in the inner crust are rather small, we expect only small changes in mode velocities if $Z$ is held fixed.

In the calculations of transverse modes,  we have included the effects of electron screening on the shear elastic constant, following the calculations of Baiko \cite{baiko}.  This leads to a reduction of the shear elastic constant by about 10\% and of the velocity of transverse modes by about 5\% for typical crustal conditions.

Mode velocities are sensitive to the neutron superfluid density, and in this connection  we have pointed to the possible importance of phonons in reducing the strength of the periodic lattice potential experienced by neutrons.

In this paper we have discussed the hydrodynamics and elasticity of the inner crust of neutron stars based on microscopic calculations of the properties of the matter \cite{lattimer} that take into account in a thermodynamically consistent way the two-phase nature of the matter.  One particularly important conclusion is that the effective neutron--proton interaction is appreciable.

There are a number of directions for future work.  One is to extend calculations to shorter wavelengths, where screening of the Coulomb interaction becomes less important.  Another is to calculate damping of modes due to either Landau damping, if electron mean free paths are long, or resistive losses, if the mean free path is short.  It is also desirable to make calculation of the effective nucleon--nucleon interactions based on state-of-the-art microscopic input, such as is obtained from chiral effective field theories of interactions between nucleons \cite{hebeler}.

We are extremely grateful to James Lattimer for providing us with output from his calculations with D. Swesty of microscopic properties of dense matter.   During the course of this work, we have had many valuable communications with Nicolas Chamel, Dany Page and Sanjay Reddy.  We have also enjoyed conversations with Dmitri Yakovlev and correspondence with Denis Baiko. Author DK acknowleges Vitaly Bychkov, Emil Lundh, and Mattias Marklund for encouragement.  This work was supported by the J.\ C.\ Kempe Memorial Fund, the Rosenfeld Foundation, and the ESF CompStar network.

\appendix\section{Variational principle}

The equation of motion may be derived in a Lagrangian formulation with the Lagrangian density
\beq
{\cal L}= -n_n{\dot\theta} + \frac12\rho^n {\dot \uu}^2- \frac12\frac{n_n^s}{m} (\nabla \theta)^2 -n_n^n {\dot\uu}\cdot\nabla \theta  \nonumber\\-E(n_n,n_p) -S\left(u_{ij}-\frac{\delta_{ij}u_{kk}}3\right)^2,
\label{Lagrangian}
\eeq
where the superfluid velocity is given by Eq.\ (\ref{vs}).  For simplicity, we use units in which $\hbar=1$ in the Appendix.  The independent variables are $\theta$, $n_n$, and the lattice displacement $\bf u$, in terms of which changes in the proton density are given by Eq.\ (\ref{deltan_p}).

Alternatively, one may use a Hamiltonian formalism: $\theta$ and $-n_n$ are conjugate variables, since the momentum conjugate to $\theta$ is $p_\theta =\partial {\cal L}/\partial{\dot \theta}=-n_n$ and the momentum density conjugate to the lattice displacement is
\beq
\pp_u=\frac{\partial {\cal L}}{\partial {\dot \uu}}=  \rho^n{\dot \uu} -n_n^n\nabla \theta
\eeq
and therefore the Hamiltonian density is
\beq
{\cal H}=\pp_u\cdot {\dot\uu} +p_\theta {\dot \theta}-{\cal L}=  \frac{\pp_u^2}{2\rho^n}+ \frac12\left(\frac{n_n^s}{m}+\frac{(n_n^n)^2}{\rho^n}\right) (\nabla \theta)^2   \,\,\,\,\,\,\nonumber\\ +\frac{n_n^n}{\rho^n} {\pp_u}\cdot\nabla \theta+E(n_n,n_p) +S\left(u_{ij}-\frac{\delta_{ij}u_{kk}}3\right)^2   \,\,\,\,\\
=  \frac{\rho_n\dot\uu^2}{2}+ \frac{n_n^s}{2m} (\nabla \theta)^2 +E(n_n,n_p) +S\left(u_{ij}-\frac{\delta_{ij}u_{kk}}3\right)^2. \nonumber\\
\label{A4}
\label{Hamiltonian}
\eeq

It is instructive to compare the formalism for a neutron superfluid moving in a lattice with that for a homogeneous mixture of neutron and proton superfluids, as in the outer core of a neutron star.  It is then natural to start from a generalized Ginzburg--Landau  approach, with a Hamiltonian density
\beq
{\cal H}^{GL}= \frac12\sum_{ij}\frac{\rho_{ij}}{m^2}\nabla \theta_i\cdot \nabla \theta_j +E(n_n,n_p),
\label{H_GL}
\eeq
where $\rho_{ij}$ is a matrix generalization of the usual superfluid stiffness coefficient or superfluid density for a one-component system (see, e.g., Ref.\ \cite{joynt}).
The current densities of the two components are then given by
\beq
\jj_i=\rho_{ij}\nabla \frac{\theta_j}{m}.
\eeq
To make contact with the results for the case of a lattice of protons, it is necessary to work in terms of the proton velocity, which may be defined in terms of the proton current density, which is given by
\beq
\jj_p=\frac{\partial{\cal H}^{GL}}{\partial \nabla\theta_p }=\frac{\rho_{pp}}{m}\nabla \theta_p+\frac{\rho_{pn}}{m}\nabla \theta_n.
\eeq
Thus the variations of the phase of the neutrons produce a gauge term in the proton current.  The mean proton  velocity is therefore
\beq
\vv_p=\frac{\rho_{pp}\nabla \theta_p+\rho_{pn}\nabla \theta_n}{m^2n_p},
\eeq
or
\beq
\nabla \theta_p=\frac{m^2n_p}{\rho_{pp}}\vv_p-\frac{\rho_{pn}}{\rho_{pp}}\nabla \theta_n.
\eeq
Thus the neutron current density is given by
\beq
\jj_n=   \left(\rho_{nn}-\frac{\rho_{np}^2}{\rho_{pp}}\right) \frac{\nabla \theta_n}{m^2}+\frac{\rho_{np}}{\rho_{pp}}n_p\vv_p.
\label{neutroncurrent2}
\eeq
By comparison of Eqs. (\ref{neutroncurrent2}) and (\ref{neutroncurrent}) one sees that the expressions are equivalent if one makes the identifications
\beq
n_n^s=\frac1m\left(\rho_{nn}-\frac{\rho_{np}^2}{\rho_{pp}}  \right)
\eeq
and
\beq
n_n^n=\frac{n_p\rho_{np}}{\rho_{pp}}.
\eeq


\begin{thebibliography}
\bibliographystyle{}
\bibitem{HaenselPotekhinYakovlev_NeutronStarsI_2007}
\mbox{P. Haensel, A. Y. Potekhin, and D. G. Yakovlev, \emph{Neutron Stars 1.}}\\ \emph{Equation of state and structure}, (Springer, New York, 2007).
\bibitem{yakovlev} D. G. Yakovlev and C. J. Pethick, {\it Ann. Rev. Nucl. Part. Sci.} {\bf 42}, 169 (2004).
\bibitem{duncan}R. C. Duncan, {\it Astrophys. J. Lett.} {\bf 498}, L45 (1998).
\bibitem{watts}T. E. Strohmayer and A. L. Watts, {\it Astrophys. J.} {\bf  653}, 593 (2006).
\bibitem{andersson} N. Andersson, K. Glampedakis, and L. Samuelsson, {\it Mon. Not. R. Astron. Soc.} {\bf  396}, 894 (2009).
\bibitem{aguilera}D. N. Aguilera, V. Cirigliano, J. A. Pons, S. Reddy, and R. Sharma, {\it Phys. Rev. Lett.} {\bf 102}, 091101 (2009).
\bibitem{epstein} R. I. Epstein,  {\it Astrophys. J.} {\bf  333}, 880 (1988).
\bibitem{carter}B. Carter and E. Chachoua, {\it Int. J. Mod. Phys. D} {\bf  15}, 1329 (2006).
\bibitem{carterChamelHaensel} B. Carter, N. Chamel, and P. Haensel, {\it Int. J.  Mod. Phys. D}
{\bf 15}, 777 (2006).
\bibitem{PethickChamelReddy_ProgrTheorPhys2010}
C. J. Pethick, N. Chamel, and S. Reddy, {\it Prog. Theor. Phys. Suppl.} {\bf 186}, 9 (2010).
\bibitem{cirigliano} V. Cirigliano, S. Reddy, and R. Sharma, {\it Phys. Rev. C} {\bf 84}, 045809 (2011).
\bibitem{ChamelPageReddy} N. Chamel, D. Page, and S. Reddy, arXiv:1210.5169.

\bibitem{lattimer} J. M. Lattimer and F. D. Swesty,  {\it Nucl. Phys. A}, {\bf 535}, 331 (1991) and the website www.astro.sunysb.edu/dswesty/lseos.html.

\bibitem{LLhydro}
L. D. Landau and E.M. Lifshitz, \emph{Course of Theoretical Physics, Vol. 6, Fluid Mechanics}, 2nd ed., (Pergamon, Oxford, 1987), \S 139.
\bibitem{baympethick}  G.\ Baym and C.\ J.\ Pethick, {\it Landau Fermi Liquid
Theory: Concepts and Applications} (Wiley, New York, 1991), Chapter 2.

\bibitem{LandauLifshitz-07}
L. D. Landau and E. M. Lifshitz, \emph{Course of Theoretical Physics, Vol. 7. Theory of Elasticity}, 3rd ed., (Pergamon, Oxford,1986).

\bibitem{kramer}  M. Kr{\"a}mer, L. Pitaevskii, and S. Stringari, {\it Phys. Rev. Lett.}
{\bf 88}, 180404 (2002).

\bibitem{shearFuchs1936}
K. Fuchs, {\it Proc. Roy. Soc. London} \textbf{153}, 622 (1936).

\bibitem{hirsekorn} S. Hirsekorn, {\it  Textures and Microstructures} {\bf 12}, 1 (1990).

\bibitem{shearNS_OgataIchimaru_1990}
S. Ogata and  S. Ichimaru, {\it Phys. Rev. A} \textbf{42}, 4867 (1990).

\bibitem{baiko} D. A. Baiko, {\it Contrib. Plasma Phys.} {\bf 52}, 157 (2012).

\bibitem{LPRL}  J. M. Lattimer, C. J. Pethick, D. G. Ravenhall, and D. Q. Lamb, {\it Nucl. Phys. A} {\bf 432} , 646 (1985).


\bibitem{Baiko2011}
D. A. Baiko,    {\it Mon. Not. Roy. Astron. Soc.} \textbf{416}, 22 (2011).

\bibitem{Chamel_BandStructCalcul_NS_2012}
N. Chamel, Phys. Rev. C \textbf{85}, 035801 (2012).

\bibitem{joynt} M. Borumand, R. Joynt, and W. Klu\'zniak,  {\it Phys. Rev. C} \textbf{54}, 2745 (1996).


\bibitem{Kittel_QuantumTheoryOfSolids}
C. Kittel, \emph{Quantum Theory of Solids}, (Wiley, New York, 1963).

\bibitem{PRE64.057402.Baiko2001b}
D. A. Baiko, A. Y. Potekhin, and D. G. Yakovlev, {\it Phys. Rev. E} {\bf 64}, 057402 (2001).

\bibitem{GnedinYakovlevPotekhin_2001}
O.Y. Gnedin, D.G. Yakovlev, and A.Y. Potekhin, {\it Mon. Not. Roy. Astron. Soc.} \textbf{324}, 725 (2001).

\bibitem{callen} See, e.g., H.\ B.\ Callen, \emph{Thermodynamics}, (Wiley, New York, 1960), Sec.\ 8.4.

\bibitem{hebeler} K.\ Hebeler and A. Schwenk, Phys.\ Rev.\ C {\bf 82}, 014314 (2010).


\end{thebibliography}
\end{document}